\documentclass[useAMS,usenatbib]{mn2e}
\usepackage{epsfig}
\usepackage{subfigure}
\usepackage{graphicx}

\def\gsim{\;\lower4pt\hbox{${\buildrel\displaystyle >\over\sim}$}\;}
\def\lsim{\;\lower4pt\hbox{${\buildrel\displaystyle <\over\sim}$}\;}
\def\grls{\;\lower4pt\hbox{${\buildrel\displaystyle >\over <}$}\;}
\begin{document}
\title[MHD Rebound Shocks in Supernovae]
{Magnetohydrodynamic Rebound Shocks of Supernovae}

\author[Y.-Q. Lou \& W.-G. Wang]{Yu-Qing Lou$^{1,2,3}$
\thanks{E-mail: louyq@tsinghua.edu.cn and lou@oddjob.uchicago.edu;
wwg03@mails.tsinghua.edu.cn } and Wei-Gang Wang$^{1}$
\\
$^{1}$Physics Department and Tsinghua Centre for Astrophysics
(THCA), Tsinghua University, Beijing, 100084, China;\\
%$^2$Centre de Physique des Particules de Marseille (CPPM)
%/Centre National de la Recherche Scientifique (CNRS)\\
%\qquad\quad
%/Institut National de Physique Nucl\'eaire
%et de Physique des Particules (IN2P3) et Universit\'e\\ \qquad\ \
%de la M\'editerran\'ee Aix-Marseille II,
%163, Avenue de Luminy, Case 902, F-13288 Marseille, Cedex 09, France;\\
$^{2}$Department of Astronomy and Astrophysics, the University
of Chicago, 5640 South Ellis Avenue, Chicago, IL 60637, USA;\\
$^{3}$National Astronomical Observatories, Chinese Academy of
Sciences, A20, Datun Road, Beijing 100012, China. \\}

%\title{MHD Rebound Shock Solutions With
%       `Quasi-Static' Asymptotes {\bf ??}}
\maketitle
%February  24,  2006 (Saturday)  THCA afternoon, evening
%February  25,  2006 (Sunday)    THCA morning, noon, afternoon, evening 
%               submitted in the evening; MN-07-0273-L
%               http://mc.manuscriptcentral.com/mnras  yqlou

\begin{abstract}
We construct magnetohydrodynamic (MHD) similarity rebound shocks 
joining `quasi-static' asymptotic solutions around the central 
degenerate core to explore an MHD model for the evolution of random 
magnetic field in supernova explosions. This provides a theoretical 
basis for further studying synchrotron diagnostics, MHD shock 
acceleration of cosmic rays, and the nature of intense magnetic 
field in compact objects. The magnetic field strength in space 
approaches a limiting ratio, that is comparable to the ratio of 
the ejecta mass driven out versus the progenitor mass, during 
this self-similar rebound MHD shock evolution. The intense 
magnetic field of the remnant compact star as compared to 
that of the progenitor star is mainly attributed to both the 
gravitational core collapse and the radial distribution of 
magnetic field.
\end{abstract}

\section{INTRODUCTION}

Self-similar evolution of a spherical gas flow under self-gravity
and thermal pressure has been studied over past four decades:
from simulations and the discovery of Larson-Penston (L-P) type
solutions (Bodenheimer \& Sweigart 1968; Larson 1969a, b; Penston
1969a, b), to the construction of the expansion-wave collapse
solution (EWCS) using the central free-fall asymptotic solution
(Shu 1977) as well as to the application of phase-match
techniques for constructing infinite series of discrete global
solutions including L-P type solutions (Hunter 1977) and 
solutions for envelope expansion with core collapse (EECC; Lou 
\& Shen 2004). Properties of eigensolutions crossing the sonic 
critical line were examined (Jordan \& Smith 1977; Shu 1977; 
Whitworth \& Summers 1985; Hunter 1986).
%; Lou \& Shen 2004).
Self-similar shocks were studied and applied to various
astrophysical settings by Tsai \& Hsu (1995), Shu et al. (2002),
Shen \& Lou (2004), and Bian \& Lou (2005). While these major
results were obtained for an isothermal gas, the counterpart
problem with a polytropic equation of state (EoS) was also studied
by Cheng (1978), Goldreich \& Weber (1980), Yahil (1983), Suto \&
Silk (1988), McLaughlin \& Pudritz (1997), Fatuzzo et al. (2004)
and Lou \& Gao (2006). In most cases, the polytropic results share
a feature that by setting the polytropic index $\gamma=1$ in the 
isothermal limit, all asymptotic behaviours approach the isothermal 
counterpart solutions. However, Lou \& Wang (2006) reported new 
`quasi-static' asymptotic solutions unique to a polytropic gas 
with $\gamma>1.2$ and constructed self-similar rebound shocks for 
supernovae (SNe).

Chiueh \& Chou (1994) studied a self-similar MHD problem by
including the magnetic pressure gradient force in the momentum
equation. Yu \& Lou (2005) improved their formulation and provided
a more detailed analysis (see Zel'dovich \& Novikov 1971 for a
discussion of random magnetic field). Wang \& Lou (2006) studied
this MHD problem for a polytropic gas and derived the `quasi-static'
asymptotic solutions. Self-similar MHD shocks were explored by Yu
et al. (2006). As magnetic field is inevitably involved in SNe and
is crucial for synchrotron radiation and cosmic ray acceleration,
we construct here rebound MHD shocks with `quasi-static' asymptotic
solutions to model magnetic field evolution in SN explosions.

Type II, Ib, Ic SNe are thought to be caused by gravitational core 
collapse due to an insufficient nuclear fuel; such collapse creates 
an over-dense core, which rebounds abruptly initiating a powerful 
rebound shock. The energetics of sustaining such a rebound shock has 
been an outstanding problem. We approach this issue in the following 
perspective.
%{\it 
Triggered by such a core collapse, the rebound shock is essentially 
supported by the neutrino-driven mechanism, and several complicated 
physical processes are involved in the stellar interior: all four 
elementary forces and the coupling of various fluids and matters 
such as baryons, neutrinos, photons etc. (e.g., Janka et al. 2006). 
We approximate such a dynamic system in terms of a single fluid 
with a polytropic EoS, and treat the shock as an energy-conserved 
self-similar shock. Conceptually, the `rebound shock' here refers to 
a neutrino-driven shock, as opposed to the `prompt shock' mentioned 
in Janka et al.  (2006).
%} 
We constructed such a rebound shock (Lou \& Wang 2006) 
to model a SN explosion followed by a self-similar 
evolution leading to a quasi-static configuration.
%which is qualitatively consistent with the rebound shock 
%scenario. We thus apply our MHD rebound shock model to 
%the physical scenario described in detail in \cite{LW06}.
In reference to the hydrodynamic model of \cite{LW06}, the main 
thrust of this Letter is to construct approximately a self-similar 
model of a quasi-spherically symmetric rebound MHD shock for a SN 
explosion, providing the profile and evolution of magnetic field 
to facilitate future studies of synchrotron radiation and MHD 
shock acceleration of cosmic rays, and to probe the nature of 
intense magnetic field of compact stellar objects left behind.
%This paper is organized as follows: Section 1 is an introduction;
%Section 2 rewrites the main theoretical formulations and solutions
%relevant to this paper; Section 3 is an analysis on magnetic fields
%of the shock break-out process; and Section 4 is the conclusion.

\section{Formulation and Analysis}

\subsection[]{The Self-Similar MHD Formulation}

A quasi-spherical similarity MHD flow embedded with a completely
random magnetic field on small scales is formulated the same as
in \cite{YuLou05} and Yu et al. (2006); the key difference here is
the polytropic EoS $p=\kappa\rho^\gamma$ instead of an isothermal
gas, where $p$ is the pressure, $\rho$ is the mass density, and
$\kappa$ is constant. Using the magnetic flux frozen-in condition,
the ideal MHD equations, viz., the mass conservation equation, the
radial momentum equation, the magnetic induction equation and the
polytropic EoS, can be reduced to two nonlinear ordinary
differential equations (ODEs)
\[
\alpha'=\alpha^{2}\bigg[(n-1)v+ \bigg(2hx +\frac
{nx-v}{3n-2}\bigg)\alpha-\frac {2(x-v)(nx-v)}{x}\bigg]
\]
\begin{equation}\label{ode1}
\qquad\times\bigg[\alpha (nx-v)^{2}
-(\gamma\alpha^{\gamma}+h\alpha^{2}x^{2})\bigg]^{-1}\ ,
\end{equation}
\[
v'=\bigg\{(n-1)\big[\alpha v(nx-v)+2h\alpha^{2}x^{2}\big]
+\frac{(nx-v)^{2}}{(3n-2)}\alpha^2
\]
\begin{equation}\label{ode2}
\qquad-2\gamma\alpha^{\gamma}\frac{(x-v)}{x}\bigg\}\bigg[\alpha
(nx-v)^{2}-(\gamma\alpha^{\gamma} +h\alpha^{2}x^{2})\bigg]^{-1}\ 
\end{equation}
along with a useful relation $m=\alpha x^2(nx-v)$ by the following 
MHD self-similar transformation in a polytropic gas flow
\[
r=k^{1/2}t^{n}x,\ \ u=k^{1/2}t^{n-1}v,\ \ 
\rho=\frac{\alpha}{4\pi Gt^{2}},\ 
\ p=\frac{kt^{2n-4}}{4\pi G}\alpha^{\gamma},
\]
\begin{equation}\label{ode3}
M=\frac{k^{3/2}t^{3n-2}}{(3n-2)G}\alpha x^2(nx-v)\ , \ \ \
<B_{t}^{2}>=\frac{kt^{2n-4}}{G}h\alpha^2x^2\ ,
\end{equation}
where $G$ is the gravitational constant, $M$ is the enclosed mass at 
time $t$ within radius $r$, $u$ is the radial flow speed, $<B_{t}^2>$ 
is the mean square of random transverse magnetic field $B_{t}$, $x$ 
is the independent self-similar variable, $v(x)$ is the reduced flow 
speed, $\alpha(x)$ is the reduced density, $m(x)$ is the reduced
enclosed mass, the prime $'$ stands for
the first derivative $d/dx$, $k$ and $n$ are two parameters, and $h$ 
is a parameter for the strength of $<B_{t}^2>^{1/2}$. We expediently
take $\gamma=2-n$ for a polytropic EoS with a constant $\kappa\equiv 
k(4\pi G)^{\gamma-1}=p\rho^{-\gamma}$. The magnetosonic critical curve 
is determined by the simultaneous vanishing of the numerator and 
denominator on the RHS of eq (1) or (2). The two eigensolutions of $v'$ 
across the magnetosonic critical curve can be derived by using the 
L'H\^ospital rule (Lou \& Wang 2006; Yu \& Lou 2005; Yu et al. 2006). 
The solutions are obtained for $v(x)$ and $\alpha(x)$, and the magnetic 
field $<B_{t}^2>^{1/2}$ is then known from transformation (\ref{ode3}).
%These two ODEs represent a subset of all possible
%solutions of the partial differential equations
%governing the quasi-spherical MHD flow,

\subsection[]{Analytic Asymptotic MHD Solutions}

For $h<h_c\equiv n^2/[2(1-n)(3n-2)]$, eqs (\ref{ode1}) and 
(\ref{ode2}) give the magnetostatic solution of a magnetized 
singular polytropic sphere (MSPS) with $v=0$ and
%\[
%v=0\ ,
%\]
\begin{equation}\label{static1}
\alpha=\bigg[\frac{n^{2}}{2\gamma(4-3\gamma)}
+\frac{(1-\gamma)}
{\gamma}h\bigg]^{-1/n}x^{{-2}/{n}}\ ,
\end{equation}
%This is a global solution with the magnetic field distribution:
\begin{equation}\label{static0}
<B_{t}^{2}>=h\frac{k^{2/n}}{G}\bigg[\frac{n^{2}}
{2\gamma(4-3\gamma)}+\frac{(1-\gamma)}
{\gamma}h\bigg]^{-2/n}r^{2-4/n}\ .
\end{equation}
There exists an asymptotic MHD solution approaching
this limiting form at small $x$ (referred to as the
type I `quasi-static' asymptotic MHD solution), viz.,
$v=Lx^K$ and
%\[
%v=Lx^K\ ,
%\]
\[
\alpha=\bigg[\frac{n^{2}}{2\gamma(4-3\gamma)}
+\frac{(1-\gamma)}{\gamma}h\bigg]^{-{1}/{n}}x^{-2/n}
\]
\begin{equation}\label{static2}
+\frac{(K+2-2/n)L}{n(K-1)}\bigg[\frac{n^2/(2\gamma)}
{4-3\gamma}+\frac{1-\gamma}
{\gamma}h\bigg]^{-1/n}x^{K-1-2/n}\ ,
\end{equation}
%\[
%v=x^{K_1}\big[L_1\cos(K_2\ln x)-L_2\sin(K_2\ln x)\big]\ ,
%\]
%\[
%\alpha=\bigg[\frac{n^{2}} {2\gamma(4-3\gamma)}+\frac{(1-\gamma)}
%{\gamma}h\bigg]^{-\frac{1} {n}}x^{\frac{-2}{n}}
%\]
%\[
%+\frac{1}{3n-2}\bigg[\frac{n^{2}}
%{2\gamma(4-3\gamma)}+\frac{(1-\gamma)} {\gamma}h\bigg]^{-\frac{1}
%{n}}x^{K_1-1-2/n}
%\]
%\begin{equation}\label{static3}
%\times\bigg\{L_1R\cos(K_2\ln x)-L_2I\sin(K_2\ln x)\bigg\}\ .
%\end{equation}
%Here for this `quasi-static' asymptotic solution,
where $K$ is the root of quadratic equation
\[
[n^2/2+n(3n-2)h]K^2-(4-3n)[n/2+(3n-2)h]K
\]
\begin{equation}\label{static4}
\qquad\qquad+n^2+\gamma(2/n-2)(3n-2)h=0\ .
\end{equation}
%$L_1$, $L_2$ and $L$ are free parameters; and:
%$$R=Re\bigg\{\bigg[\big(\frac{n^2}
%{2(3n-2)}+nh\big)(K_1+K_2i)+\frac{n^2}{2(3n-2)}-\gamma
%h\bigg]^{-1}\bigg\}\ ,$$
%$$I=Im\bigg\{\bigg[\big(\frac{n^2}{2(3n-2)}
%+nh\big)(K_1+K_2i)+\frac{n^2}{2(3n-2)}-\gamma
%h\bigg]^{-1}\bigg\}\ ,$$respectively. Note that
%We define $h_0\equiv\big[(3+2\sqrt{2})n-4\big]
%\big[4-(3-2\sqrt{2})n\big]/[2n(3n-2)]$.
When $12-8\sqrt 2<n<0.8$ and $h_0<h<h_c$,
where $h_0\equiv\big[(3+2\sqrt{2})n-4\big]
\big[4-(3-2\sqrt{2})n\big]/[2n(3n-2)]$, or when
$2/3<n<12-8\sqrt{2}$ for $h<h_c$, eq (\ref{static4})
gives two roots $K>1$, corresponding to 
two possible `quasi-static' solutions.
%while when $12-8\sqrt{2}<n<0.8$ and
%$h<h_0$, there exists a complex root
%$K$ with real parts larger than 1.

The asymptotic MHD solution at large
$x$ is $\alpha=A_0x^{-2/n}+\cdots$ and
\[
v=B_0x^{1-1/n}-\bigg[\frac{n}{(3n-2)}
+\frac{2h(n-1)}{n}\bigg]A_0x^{1-2/n}
\]
\begin{equation}\label{large1}
\qquad+\frac{2\gamma A_0^{\gamma-1}}
{n[2(n+\gamma)-3]}x^{(2-2\gamma-n)/n}+\cdots\ ,
\end{equation}
%\begin{equation}\label{large2}
%\alpha=A_0x^{-2/n}\ ,
%\end{equation}
where $A_0$ and $B_0$ are two 
constants. Solution (\ref{large1})
%and (\ref{large2})
at large $x$ can be connected to `quasi-static'
asymptotic MHD solution (\ref{static2}) and 
(\ref{static4}) at small $x$ by a Runge-Kutta 
integration (Press et al. 1986), crossing the 
magnetosonic critical curve either smoothly or 
with an MHD shock (Yu et al. 2006).
%\cite{LW06}.

\subsection[]{MHD Shock Jump Conditions}

MHD shock conditions (Yu et al. 2006; Lou \& Wang 2006)
%[\cite{YuLou06} and \cite{LW06}]
%for shocks in the isothermal MHD case and in
%the polytropic hydrodynamic case, respectively]
include conservations of mass, momentum, energy and
magnetic flux, and in self-similar forms, they appear as
\begin{equation}\label{shock1}
\bigg[\alpha_{s}\bigg(n-\frac{v_{s}}{x_{s}}\bigg)\bigg]^1_2=0\ ,
\end{equation}
%\begin{equation}\label{shock1}
%\alpha_{s1}\bigg(n-\frac{v_{s1}}{x_{s1}}\bigg)
%=\alpha_{s2}\bigg(n-\frac{v_{s2}}{x_{s2}}\bigg)\ ,
%\end{equation}
%
\begin{equation}\label{shock2}
\bigg[\frac{\alpha_{s}^{\gamma}}{x_{s}^2}
+\alpha_{s}\bigg(n-\frac{v_{s}}{x_{s}}\bigg)^2
+\frac{h\alpha_{s}^2}{2}\bigg]^1_2=0\ ,
\end{equation}
%\begin{equation}\label{shock2}
%\frac{\alpha_{s1}^{\gamma}}{x_{s1}^2}
%+\alpha_{s1}\bigg(n-\frac{v_{s1}}{x_{s1}}\bigg)^2
%+\frac{h\alpha_{s1}^2}{2}
%=\frac{\alpha_{s2}^{\gamma}}{x_{s2}^2}
%+\alpha_{s2}\bigg(n-\frac{v_{s2}}{x_{s2}}\bigg)^2
%+\frac{h\alpha_{s2}^2}{2}\ ,
%\end{equation}
%
\begin{equation}\label{shock3}
\bigg[\bigg(n-\frac{v_{s}}{x_{s}}\bigg)^2\!\!\!
+\frac{2\gamma}{(\gamma-1)}\frac{\alpha_{s}^{\gamma-1}}
{x_{s}^2}+2h\alpha_{s}\bigg]^1_2=0\ ,
\end{equation}
%\begin{equation}\label{shock3}
%\bigg(n-\frac{v_{s1}}{x_{s1}}\bigg)^2\!\!\!
%+\frac{2\gamma}{\gamma-1}\frac{\alpha_{s1}^{\gamma-1}}
%{x_{s1}^2}+2h\alpha_{s1}=\bigg(n-\frac{v_{s2}}{x_{s2}}
%\bigg)^2\!\!\!+\frac{2\gamma}{\gamma-1}
%\frac{\alpha_{s2}^{\gamma-1}}{x_{s2}^2}+2h\alpha_{s2}\ ,
%\end{equation}
where quantities in square brackets with superscript `1'
(upstream) and subscript `2' (downstream) remain conserved
across the MHD shock front indicated by a subscript $_s$.
%where quantities with a subscript `s1' denotes the quantities
%at the upstream side of the shock and quantities with a
%subscript `s2' denotes those at the downstream side.
The parameter $k$ changes according to $k_2=k_1x_{s1}^2/x_{s2}^2$
on two sides of a shock. For the specific entropy to increase from
upstream to downstream sides, $x_{s1}>x_{s2}$ is necessary. MHD
shock conditions (\ref{shock1})$-$(\ref{shock3}) lead to a 
quadratic equation (Lou \& Wang 2006); once we specify physical 
conditions on one side of a chosen shock location, the corresponding 
quantities $\alpha,\ v,\ x$ on the other side are readily computed.

\section[]{Rebound MHD Shocks in Supernova Explosions}

Various rebound MHD shocks are constructed numerically, parallel
to \cite{LW06}. With chosen inner and outer radii, e.g.,
$r_{\mbox{i}}=10^6$cm and $r_{\mbox{o}}=10^{12}$cm for neutron
star formation, and when the $k$ parameter in transformation
(\ref{ode3}) is specified, we apply our solutions to a physical 
rebound MHD shock scenario for SNe (Lou \& Wang 2006).

\subsection[]{Final and Initial Configurations}

Similar to the hydrodynamic rebound shock model of Lou \& Wang (2006), 
the final configuration (small $x$) of our rebound MHD shock solutions 
gradually evolves to a MSPS and is regarded as a remnant compact object 
after the rebound MHD shock ploughing through stellar ejecta; the 
initial configuration (large $x$) marks the onset of gravity-induced 
core collapse with outer inflows or outflows such as stellar winds or 
stellar oscillations.

We define the outer initial mass $M_{\mbox{o,ini}}$ and 
the inner ultimate mass $M_{\mbox{i,ult}}$ the same way 
as in \cite{LW06} and regard them as rough estimates for 
the masses of the progenitor star and the remnant 
compact object. The ratio of the two masses is
$M_{\mbox{o,ini}}/M_{\mbox{i,ult}}
=\lambda_1(r_{\mbox{o}}/r_{\mbox{i}})^{(3-2/n)}$
%\begin{equation}\label{ratio1}
%M_{\mbox{o,ini}}/M_{\mbox{i,ult}}
%=\lambda_1(r_{\mbox{o}}/r_{\mbox{i}})^{(3-2/n)}\ ,
%\end{equation}
where $\lambda_1\equiv A_0(k_1/k_2)^{1/n}
\big\{n^{2}/[2\gamma(4-3\gamma)]
+(1-\gamma)h/\gamma\big\}^{1/n}$ involves parameters
of the rebound MHD shock and is equal to the ratio of
enclosed masses at the same $r$.
%(or take $r_{\mbox{o}}=r_{\mbox{i}}$ in this case).
Similar to the result of \cite{LW06}, we find numerically
that $\lambda_1>1$ depends on the choice of solutions,
clearly indicating that a rebound MHD shock drives out
stellar materials.

By eq (\ref{ode3}),
%the magnetic field strength scales as $B_t\propto r^{1-2/n}$.
%Moreover, the initial and final configurations are both static.
the final magnetostatic configuration gives
\[
<B_{t,\mbox{ult}}^2>^{1/2} =\sqrt{\frac{h}{G}}\bigg[\frac{n^{2}}
{k2\gamma(4-3\gamma)}+\frac{1-\gamma}
{k\gamma}h\bigg]^{-1/n}r^{1-2/n}\ .
\]
\begin{equation}\label{ratio2}
\\
\end{equation}
The ratio of initial to final magnetic fields at the same $r$ is
$<B_{t,\mbox{ini}}^2>^{1/2}/<B_{t,\mbox{ult}}^2>^{1/2}=\lambda_1$,
%\begin{equation}\label{ratio3}
%\frac{\sqrt{<B_{t,\mbox{ini}}^2>}}
%{\sqrt{<B_{t,\mbox{ult}}^2>}}=\lambda_1\ ,
%\end{equation}
where $\lambda_1>1$ by numerical exploration.
%is larger than, but not significantly deviated from, unity.
Thus a rebound MHD shock breakout process {\it reduces} the magnetic
field by the same ratio of enclosed masses at the same $r$; yet this
decrease in magnetic field is insignificant as compared to the radial
variation of magnetic field, i.e., the $r^{1-2/n}$ dependence. As
$\gamma$ approaches $4/3$ or $n\rightarrow 2/3$, this scaling
approaches $r^{-2}$, while the dependence of enclosed mass on $r$ 
goes to $r^0$. For a $\sim 10$G (0.1G) surface magnetic field at
%the initial outer boundary, i.e., at the stellar surface (
$r_{\mbox{o}}=10^{12}$cm, we estimate a magnetic field in the
interior of the final configuration ($r_{\mbox{i}}=10^6$cm) to be
$\sim 10^{13}$G ($10^{11}$G), sensible for magnetized neutron
stars; if we take $r_{\mbox{i}}=10^9$cm, then the final interior
magnetic field is estimated to be $\sim 10^7$G ($10^5$G), fairly
close to relevant magnetic field strengths of white dwarfs (e.g.,
Euchner et al. 2005, 2006; Schmidt et al. 2003).

\subsection[]{Evolution of Rebound MHD Shocks}

Time evolution of density, velocity and enclosed mass are similar
to those described by \cite{LW06}. We focus here on the magnetic
field evolution. Figure 1 shows a typical time evolution of
$<B_{t}^2>^{1/2}/<B_{t,\mbox{ult}}^2>^{1/2}$ to complement the
$r^{1-2/n}$ behaviour. Magnetic field increases at first, and
gradually decreases until reaching the magnetostatic configuration
much smaller than the initial configuration in size. In short, 
magnetic field changes moderately.
%From the above numerical results, we see that the magnetic
%field does not change significantly (although decreases
%slightly) during the process of rebound shock breakout, yet
The crucial point is that the magnetic field varies significantly
in $r$ within a star. If we take the magnetic field at the outer
boundary to be the surface magnetic field of the progenitor star
and take the magnetic field at the inner boundary as the surface
magnetic field of the remnant compact star, then a large ratio of
$\sim 10^{12}$ appears in forming a neutron star (Lou \& Wang
2006). This model feature may explain the intense magnetic field
of neutron stars inferred from spin-down observations of radio
pulsars. In our scenario, after the passage of such a rebound MHD
shock, stellar ejecta detach from the central degenerate neutron 
star which is thus exposed with a surface magnetic field of 
$10^{13\sim 11}$G. In the same spirit of Lou \& Wang (2006), we 
also suggest the formation of magnetic white dwarfs from the end 
of main-sequence stars with $6\sim 8M_{\odot}$; in this scenario, 
the surface magnetic field of an exposed central white dwarf is 
in a plausible range of $\sim10^{7\sim5}$G (e.g., Euchner et al. 
2005, 2006; Schmidt et al. 2003).

\begin{figure}
\includegraphics[width=3.3in,bb=120 270 480 570]{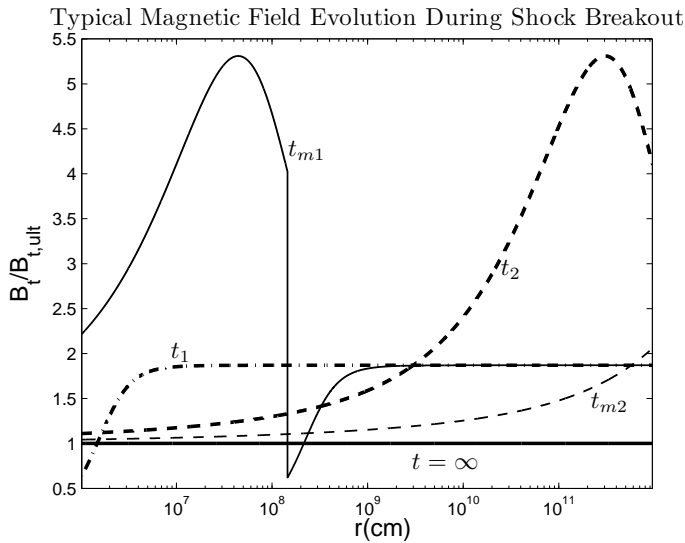}
\caption{The ratio $B_{t}/B_{t,\mbox{ult}}\equiv
<B_{t}^2>^{1/2}/<B_{t,\mbox{ult}}^2>^{1/2}$ is the rms magnetic
field strength divided by the corresponding rms magnetic field
strength of the final magnetostatic configuration at the same $r$.
This example is constructed by integrating inward from $(x_0\
,v_0\ ,\alpha_0)$ on the magnetosonic critical curve and using an
eigensolution to match with a quasi-static solution as
$x\rightarrow 0^{+}$; we use the solution portion within
$x_{s2}<x_0$ for the downstream. We then obtain the upstream point
$(x_{s1} ,v_{s1} ,\alpha_{s1})$ by the MHD shock jump conditions
from the values of $(x_{s2} ,v_{s2} , \alpha_{s2})$ obtained in
the former integration and further integrate outward to determine
the upstream solution. The relevant parameters are $\gamma=1.32$,
$n=0.68$, $h=0.01$, $k_1=7.7\times10^{16}$ cgs units,
$k_2=4\times10^{17}$ cgs units, $x_0=1.778\ , v_0=0.4620\ ,
\alpha_0=0.067$, and $x_{s_2}=1.1$.
%(see Lou \& Wang 2006 for specific steps).
Here, $t_1=6.61\times 10^{-5}$s is the time when the MHD shock
crosses the inner boundary and is the initial time of application;
$t_2=4.40\times 10^4$s is the time when the MHD shock crosses the
outer boundary; $t_{m1}=0.1$s and $t_{m2}=1\times10^8$s are two
intermediate times between $t_1$ and $t_2$ and $t_2$ and
$t=\infty$.}\label{MagEvol}
\end{figure}

The major point to be emphasized is that random magnetic field 
preexists inside progenitor stars through various dynamo processes.
We detect magnetic field strengths of order $10^{-2\sim 3}$G on
stellar surface and this corresponds to a much stronger magnetic
field in the stellar interior with a scaling of $\sim r^{1-2/n}$
shortly after the initiation of core collapse. In addition, the 
interior magnetic field can be considerably strengthened by the
free-fall core collapse preceding the emergence of a rebound MHD 
shock (see Lou \& Wang 2006 for descriptions of the rebound shock 
scenario and the core collapse process), according to the frozen-in 
flux and accretion shock conditions. In reality, these two processes
happen concurrently to produce the resultant self-similar distribution 
of magnetic field. In short, the interior magnetic field would be 
much stronger than the surface magnetic field and can be further 
enhanced to reach a high-field regime.

The origin of stellar magnetic field was argued by several authors 
to come from various processes, including dynamo effects and 
thermomagnetic instabilities (e.g., Reisenegger et al. 2005). Our 
MHD scenario of interior core collapse and rebound shock appears 
to grossly match with observational facts.
%Although a quantitative analysis is still needed to justify this 
%mechanism, we can still provide a qualitative justification of the 
%large interior magnetic field involved in the proposed mechanism.
From our MSPS configuration with a random magnetic field strength 
scaled as $B_t\propto r^{1-2/n}$ in a polytropic gas, we see a 
real possibility that the interior magnetic field can be actually 
much stronger than the surface magnetic field. Once the onset of 
a gravitational collapse has been initiated within a magnetized 
progenitor star and following subsequent free-fall core collapse 
and accretion shock, an eventual emergence of a rebound MHD shock 
can evolve in a quasi-spherical self-similar manner and can end 
up to a MSPS configuration with a high-density compact degenerate 
object left behind.

\section[]{Conclusions and Discussion}

We outline and propose the model scenario of a quasi-spherical rebound 
MHD shock to form high-density compact stars after a gravity-induced
collapse in the core of a progenitor star that runs out of nuclear 
fuels. The stellar interior magnetic field is expected to be enhanced
during the core collapse before the eventual emergence of a rebound 
MHD shock; also the interior magnetic field should be much stronger 
than the stellar surface magnetic field prior to the onset of a core 
collapse and during the outward propagation of the rebound MHD shock. 
Once the magnetostatic configuration of a remnant degenerate star 
appears, stellar ejecta gradually detach from the compact object, 
exposing intense surface magnetic fields of $\sim 10^{13\sim 11}$G 
for neutron stars or $\sim 10^{7\sim 5}$G for magnetic white dwarfs.

Formally, MSPS solution (\ref{static1}) for density diverges as 
$x\rightarrow 0^+$. Conceptually, this can be readily reconciled 
by the onset of degeneracy in core materials at a nuclear mass 
density. 
%Likewise, solution (\ref{large1})

%{\it 
In our model, there are two parameters for magnetic field: index 
$n$ for radial variation and ratio $h$. While it appears in this Letter 
that $n$ depends on the stiffness (i.e., $\gamma$) of EoS, as discussed 
below it is in fact a parameter free from the stiffness (i.e., $\gamma$). 
Meanwhile, ratio $h$ represents an ideal MHD approximation that dictates 
the magnitude variation of random transverse magnetic field; other 
factors, such as metalicity, differential rotation, convective motions, 
buoyancy etc. (Janka 2006), are important in generating random magnetic 
fields inside a star prior to the onset of the core collapse. 
%It is believed that magnetic field strengths are related to 
%(differential) rotation, yet our model mainly discusses the 
%large-scale behaviour of this rotation-generated magnetic field, and 
%this behaviour is assumed to be more or less irrelevant to the 
%large-scale rotation. Hence a quasi spherical symmetry is assumed here. 
%}

In contexts of SN explosions, two-shock models, i.e., models
involving a `forward shock' for the SN remnant shock after the
powerful rebound shock crashing into the interstellar medium
and a `reverse shock' produced by the same impact process (see,
e.g., Chevalier et al.
%Blondin \& Emmering
1992 and Truelove \& McKee 1999), have been studied earlier.
The major formulation difference between these earlier works
and ours is that they ignored self-gravity of stellar ejecta.
%This difference accounts for the fact that we cannot
%include a two-shock model (with one going forward and
%one backward) in our self-similar framework.
By estimates, the self-gravity cannot be obviously dropped
and thus these models including forward and reverse shocks
would be applicable in the limit of extremely strong shocks
in order to ignore self-gravity.
%can be discarded or be considered as some kind of
%approximation (the existence of two-shock systems
%are certainly approved by observations).
Another major difference is that these earlier models focus
on circumstellar interactions, while we focus on a rebound
MHD shock as it travels within the magnetized stellar interior.

Our polytropic model is currently restricted to $\gamma=2-n$ for 
a constant $\kappa$ merely for expediency. This constraint can be 
actually removed if we consistently allow the reduced pressure to 
be $\propto\alpha^{\gamma}m^q$ where index parameter $q\equiv 
2(n+\gamma-2)/(3n-2)\neq 0$ in general and $m=\alpha x^2(nx-v)$ is 
the reduced enclosed mass. It is then possible for $1<\gamma<2$ 
while $n\rightarrow 2/3$. This more general case will be reported 
separately (Wang \& Lou 2006).

Numerical MHD simulations and observations are needed to further
test our scenario for rebound MHD shocks in SNe, such as direct 
or indirect observation of density and flow speed profiles (Lou 
\& Wang 2006) as well as diagnostics of synchrotron emissions 
caused by relativistic electrons in random magnetic field 
generated by MHD shocks.

%We note here that our MHD model is established upon the dominance
%of radial flow, yet a static or `quasi-static' %configuration
%violates this condition. Thus even if perfect quasi-spherical
%MHD flow survives earlier evolutions of the system, turbulent
%flows may ruin the symmetry completely at the end.

\section*{Acknowledgments}
This research has been supported in part by the ASCI Center
for Astrophysical Thermonuclear Flashes at the Univ. of Chicago,
%by the Special Funds for Major State Basic
%Science Research Projects of China,
by THCA,
%by the Tsinghua Center for Astrophysics,
%by the Collaborative Research Fund from the National Science
%Foundation of China (NSFC) for Young Outstanding Overseas
%Chinese Scholars (NSFC 10028306) at the National Astronomical
%Observatories, Chinese Academy of Sciences,
by the NSFC grants 10373009 and 10533020 at the Tsinghua
Univ., and by the SRFDP 20050003088 and the Yangtze Endowment
from the Ministry of Education at the Tsinghua Univ.
%Affiliated institutions of Y-QL share this contribution.

\end{document}